\RequirePackage[pagewise,mathlines]{lineno}
\documentclass[twocolumn,showpacs,amsmath,amssymb,superscriptaddress,nofootinbib]{revtex4-1}

\usepackage{graphicx}
\usepackage{dcolumn}
\usepackage{bm}
\usepackage{color}
\usepackage{ulem}
\usepackage{url}
\usepackage{multirow}
\usepackage{enumitem}

\newcommand{\bfg }{\begin{figure}[htpb]}
\newcommand{\efg }{\end{figure}}
\newcommand{\bmn }{\begin{minipage}}
\newcommand{\emn }{\end{minipage}}
\newcommand{\bt }{\begin{table}[htpb]}
\newcommand{\et }{\end{table}}

\newcommand{\sNN}{$\sqrt{s_{\rm NN}}$ }

\newcommand{\GeVc}{GeV/$c$ }

\newcommand{ \be }{\begin{equation}}
\newcommand{ \ee }{\end{equation}}
\newcommand{ \bea }{\begin{eqnarray}}
\newcommand{ \eea }{\end{eqnarray}}

\newcommand{ \pT}{$p_{\rm T}$}

\newcommand {\pt}	{p_{T}}
\newcommand {\ptrho}	{p_{T,\rho}}
\newcommand {\vexe}	{v_{2,{\rm ebye}}}
\newcommand {\vobs}	{\vexe^{\rm obs}}
\newcommand {\vrho}	{v_{2,\rho}}
\newcommand {\vrhoexe}	{v_{2,\rho,{\rm ebye}}}
\newcommand {\vrhoobs}	{v^{\rm obs}_{2,\rho,{\rm ebye}}}

\newcommand {\vpiexe}	{v_{2,\pi,{\rm ebye}}}
\newcommand {\vpiobs}	{v^{\rm obs}_{2,\pi,{\rm ebye}}}

\newcommand {\psiPP}	{\psi_{\rm PP}}
\newcommand {\psiEP}	{\psi_{\rm EP}}
\newcommand {\phires}	{\phi_{\rm res}}
\newcommand {\phirho}	{\phi_{\rho}}
\newcommand {\vres}	{v_{2,{\rm res}}}
\newcommand {\Bvec}	{\vec{B}}

\newcommand {\gSS}	{\gamma_{\rm SS}}
\newcommand {\gOS}	{\gamma_{\rm OS}}
\newcommand {\gdel}	{\Delta\gamma}
\newcommand {\piinc}	{\pi_{\rm inc}}

\newcommand {\mean}[1]	{\langle #1\rangle}

\begin{document}
\def\Journal#1#2#3#4{{#1} {\bf #2}, #3 (#4)}

\def\NCA{Nuovo Cimento}
\def\NIM{Nucl. Instr. Meth.}
\def\NIMA{{Nucl. Instr. Meth.} A}
\def\NPB{{Nucl. Phys.} B}
\def\NPA{{Nucl. Phys.} A}
\def\PLB{{Phys. Lett.}  B}
\def\PRL{{Phys. Rev. Lett.}}
\def\PRC{{Phys. Rev.} C}
\def\PRD{{Phys. Rev.} D}
\def\ZPC{{Z. Phys.} C}
\def\JPG{{J. Phys.} G}
\def\EPJ{{Eur. Phys. J.} C}
\def\EPJA{{Eur. Phys. J.} A}
\def\EPJST{{Eur. Phys. J.} - Special Topics}
\def\RPP{{Rep. Prog. Phys.}}
\def\PR{{Phys. Rep.}}
\def\ANP{{Adv. Nucl. Phys.}}

\preprint{}
\bibliographystyle{h-physrev}


\title{Challenges in flow background removal in search for the chiral magnetic effect}

\author{Fuqiang Wang}
\affiliation{School of Science, Huzhou University, Huzhou, Zhejiang 313000, China}
\address{Department of Physics and Astronomy, Purdue University, West Lafayette, Indiana 47907, USA}
\author{Jie Zhao}
\email{zhao656@purdue.edu}
\address{Department of Physics and Astronomy, Purdue University, West Lafayette, Indiana 47907, USA}


\begin{abstract}

We investigate the effect of resonance decays on the three-particle correlator charge separation observable in search for the chiral magnetic effect, using a simple simulation with realistic inputs. We find that resonance decays can largely account for the measured signal. We suppress the elliptic flow ($v_2$) background by using zero event-by-event $v_2$ (or via the mixed-event technique). We find that the background is suppressed, but not eliminated as naively anticipated. We identify the reason to be the non-identicalness of the resonance and final-state particle's $v_2$ and the induced correlation between the transverse momentum dependent resonance $v_2$ and decay angle. We make predictions for the charge separation signal due to resonance decays in 200~GeV Au+Au collisions. 
\end{abstract}

\maketitle



{\it Introduction.}
Metastable domains of deconfined quark matter may form in quantum chromodynamics (QCD) where the topological charge can fluctuate to non-zero values
~\cite{Kharzeev:2015znc}.
Interactions with the topological charge field would change the overall quark chirality in those domains where the approximate chiral symmetry may be restored. These phenomena could arise in relativistic heavy ion collisions~\cite{Arsene:2004fa,Back:2004je,Adams:2005dq,Adcox:2004mh}. The strong magnetic field produced by the spectator protons in those collisions would then induce an electric current of the chirality imbalanced quark matter, resulting in charge separation of final-state particles--the chiral magnetic effect (CME)~\cite{Kharzeev:2007tn}. An observation of the CME-induced charge separation would confirm a fundamental property of QCD and is therefore of paramount importance.

Because of the fluctuation nature of the finite topological charge, the induced charge separation can only be measured by particle correlation method. 
One proposed observable~\cite{Voloshin:2004vk} is the three-particle correlator $\cos(\alpha+\beta-2c)$ where $\alpha$, $\beta$, 
and $c$ are the azimuthal angles of three particles. The particle $c$, neglecting all nonflow effects and after resolution correction, 
is a surrogate of the participant plane $\psiPP$~\cite{Alver:2006wh,PPRP:2017}.
\begin{equation}
\gamma\equiv\cos(\alpha+\beta-2\psiPP)\,.
\label{eq:C}
\end{equation}
Charge separation along the magnetic field ($\Bvec$) perpendicular to $\psiPP$ on average, would yield different values of $\gamma$ for particle pairs of same-sign (SS) and opposite-sign (OS) charges: $\gdel\equiv\gOS-\gSS>0$. A positive $\gdel$, referred to as the charge separation signal, would therefore signal existence of the CME.

There is, however, mundane physics that differ between SS and OS pairs~\cite{Wang:2009kd,Bzdak:2009fc,Liao:2010nv,Pratt:2010zn,Schlichting:2010qia}. 
One such physics is resonance decays, such as $\rho^0\rightarrow\pi^+\pi^-$. Because of resonance elliptic anisotropy ($\vres$), 
more OS pairs align in the $\psiPP$ than $\Bvec$ direction, an anti-charge separation along $\psiPP$. 
This would mimic the same effect as the CME on the $\gdel$ variable~\cite{Voloshin:2004vk,Abelev:2009ac,Abelev:2009ad}. This flow background is 
\begin{eqnarray}
\gdel&\propto&\mean{\cos(\alpha+\beta-2\phires)\cos2(\phires-\psiPP)}\nonumber\\
&\approx&\mean{\cos(\alpha+\beta-2\phires)}\vres\,.
\label{eq:bkgd}
\end{eqnarray}
Flow anisotropy can be measured by the Q-method: $\vobs = Q^{*}q_{EP}$, 
where $Q=\frac{1}{N}\sum_{j=1}^{N} e^{2i\phi_j}$ summing over particles (i.e.~those for $\alpha$ and $\beta$) in each event, 
and $q_{EP}=e^{2i\psiEP}$; $\psiEP$ is called the event plane (EP), 
reconstructed from final-state particles, as a proxy for $\psiPP$ that is not experimentally accessible. 
Particles used for $Q$ and EP calculations are exclusive to each other; one usually divide the event into two sub-events, 
one for $Q$ and the other for EP.
To suppress the flow background, STAR has applied the event-by-event method and extracted the charge separation 
signal\footnote{What was measured in the STAR work~\cite{Adamczyk:2013kcb} is not identical to $\gdel$ 
but closely related.} at $\vobs=0$~\cite{Adamczyk:2013kcb}. A recent study~\cite{Wen:2016zic} proposes a tighter cut, $|Q|=0$, 
to extract signal. 
It requires, however, small-Q extrapolation where the signal dependence is not obvious and, 
because of the zero phase space at $|Q|=0$, may suffer from large uncertainties.

\begin{table*}
\caption{Simulation inputs: primordial $\pi^{\pm}$ rapidity densities $dN_{\pi^{\pm}}/dy$ (obtained from inclusive pion $dN/dy$ minus resonance contributions, and assumed $\pi^+=\pi^-$), and $\pt$ spectra $dN_{\pi^{\pm}}/dm^2_{T}\propto(e^{m_{T}/T_{BE}}-1)^{-1}$ where $m_T=\sqrt{\pt^2+m_{\pi}^2}$ ($m_{\pi}$ is the $\pi^{\pm}$ rest mass); $dN/dy$ ratios of resonances to inclusive pion ($\piinc\equiv\piinc^++\piinc^-$), assumed centrality independent, and $\rho$ $\pt$ spectrum (obtained from fit to 200~GeV Au+Au data of the 40-80\% centrality~\cite{Adams:2003cc}) used for all resonances ($\rho,\eta,\omega$) in all centralities; and $v_2/n=a/(1+e^{-[(m_{T}-m_{0})/n-b]/c})-d$, where $n=2$ is the number of constituent quarks (NCQ) and $m_0$ is the particle rest mass for $\pi, \rho, \eta, \omega$, respectively. The $T_{BE}$ and $\piinc$ $dN/dy$ are from Bose-Einstein fit to the measured inclusive pion spectra~\cite{Adler:2003qi,Adams:2003xp}, and the $a,b,c,d$ parameters are from fit to the measured inclusive pion $v_2$~\cite{Adams:2004bi,Adare:2010sp} by the NCQ-inspired function~\cite{Dong:2004ve}.}
\centering
\begin{tabular}{l|cccccc|c}
\hline
Centrality  &  $dN_{\pi^{\pm}}/dy$ & $T_{BE}$ (GeV) & $a$ & $b$ (GeV) & $c$ (GeV) & $d$ & Resonances $\rho,\eta,\omega$ \\  \hline
70-80\% & 7.8  & 0.171 & 0.118 & 0.180 & 0.155 & 0.024 & \\
60-70\% & 16.7 & 0.179 & 0.140 & 0.116 & 0.173 & 0.046 & $dN/dy$ ratios: \\
50-60\% & 31.9 & 0.185 & 0.123 & 0.157 & 0.155 & 0.029 & $2\rho/\piinc=0.169$~\cite{Adams:2003cc}, \\
40-50\% & 53.9 & 0.190 & 0.136 & 0.145 & 0.175 & 0.039 & $\eta/\rho=0.47$, $\omega/\rho=0.59$~\cite{Adamczyk:2015lme} \\
30-40\% & 85.7 & 0.195 & 0.125 & 0.170 & 0.177 & 0.031 & $\pt$ spectra: \\
20-30\% & 129  & 0.198 & 0.125 & 0.147 & 0.210 & 0.039 & $\frac{d^2N_{\rm res}}{m_Tdm_Tdy}=\frac{dN_{\rm res}/dy}{T(m_{0}+T)}e^{-(m_{T}-m_{0})/T}$ \\
10-20\% & 186  & 0.219 & 0.096 & 0.155 & 0.212 & 0.030 & $T=0.317$~GeV~\cite{Adams:2003cc} \\
 0-10\% & 262  & 0.219 & 0.041 & 0.214 & 0.145 & 0.006 & \\
\hline
\end{tabular}
\label{datatab}
\end{table*}

Experimentally, it is very challenging, if not at all impossible, to measure the event-by-event $\vobs$ of resonances such as the $\rho$. One instead uses the $\vobs$ of final-state particles as was done in Ref.~\cite{Adamczyk:2013kcb}. Furthermore, the background in Eq.~(\ref{eq:bkgd}) is proportional to $\vres$ only when $\cos(\alpha+\beta-2\phires)$ and $\cos2(\phires-\psiPP)$ can be factorized. This may not be the case because both quantities depend on the transverse momentum ($\pt$) of the resonance. In this paper, we investigate the effects of these approximations on the premise of flow background suppression in the $\gdel$ observable, by using {\it Monte Carlo} (MC) simulation of resonance decays with realistic inputs from data for the resonance kinematic distributions. 

{\it Simulation setup.}
We focus on charged pion correlations in our study. We generate primordial pions and resonances that decay into pions according to centralities in Au+Au collisions at 200~GeV.
We include $\rho \rightarrow \pi^{+}\pi^{-}$ (branching ratio $\sim100\%$), $\eta \rightarrow \pi^{+}\pi^{-}\pi^{0}$ ($\sim22.9\%$), $\eta \rightarrow \pi^{+}\pi^{-}\gamma$ ($\sim4.2\%$), $\omega \rightarrow \pi^{+}\pi^{-}\pi^{0}$ ($\sim89.2\%$), and $\omega \rightarrow \pi^{+}\pi^{-}$ ($\sim1.5\%$)~\cite{Agashe:2014kda}. 
The inputs to the simulation 
are listed in Table~\ref{datatab}. A few notes: 
%
%
%
(i) Listed in Table~\ref{datatab} are Bose-Einstein fit parameters to the inclusive pion ($\piinc$) spectra ~\cite{Adler:2003qi,Abelev:2008ab}, not primordial,
so our spectra are somewhat off. 
(ii) We fit the combined $\piinc$ $v_2$ data from STAR~\cite{Adams:2004bi} and PHENIX~\cite{Adare:2010sp} to a function inspired by the number of constituent quarks (NCQ) scaling~\cite{Dong:2004ve}. We apply those $v_2$ to our primordial pions, so our pion $v_2$ is also somewhat off. 
In practice, we input a $\overline{v_{2}}$ and a $v_{2}$ dynamical fluctuation of $\sigma=40\%\overline{v_{2}}$ 
into our simulation, such that $v_{2}\{2\} = \sqrt{\overline{v_{2}}^{2}+\sigma^{2}}$ is given by the above NCQ-scaling inspired fit to the experimental data.
(iii) 
We use the same spectral shape for all resonances in all centralities. 

We use realistic input to the MC simulation as much as we can. However, experimentally there are uncertainties on these inputs, 
many of which (e.g. the resonance $\pt$ and $v_{2}$ spectrum) are not measured except the $\rho$ $\pt$ spectrum in $40-80\%$ centrality ~\cite{Adams:2003cc}. 
We apply our best understanding of the available experimental data to the simulation input to assess 
how much effect resonance decays can cause. 
Because of these uncertainties, our results should be taken semi-qualitatively in comparison to experimental data, 
perhaps within a factor of two. However, our qualitative conclusions regarding the large resonance contributions 
to the $\gdel$ correlator is important.

In our simulation, we take flat $dN/dy$ distributions in $|y|<1.5$ for $\pi^{\pm}$ $\eta$, $\omega$ and $\rho$. 
For each event, we sample the particle multiplicities by Poisson statistics.
Resonance mass distributions are sampled according to Breit-Wigner function with the mass and width from PDG~\cite{Agashe:2014kda}. 
The two and three-body phace-space decay methods are used. We use the final $\pi$ within $\pt > 0.2$~\GeVc and $|\eta|<1$ (i.e.~the STAR acceptance) in our analysis.

\begin{figure}\centering
\includegraphics[width=0.49\textwidth]{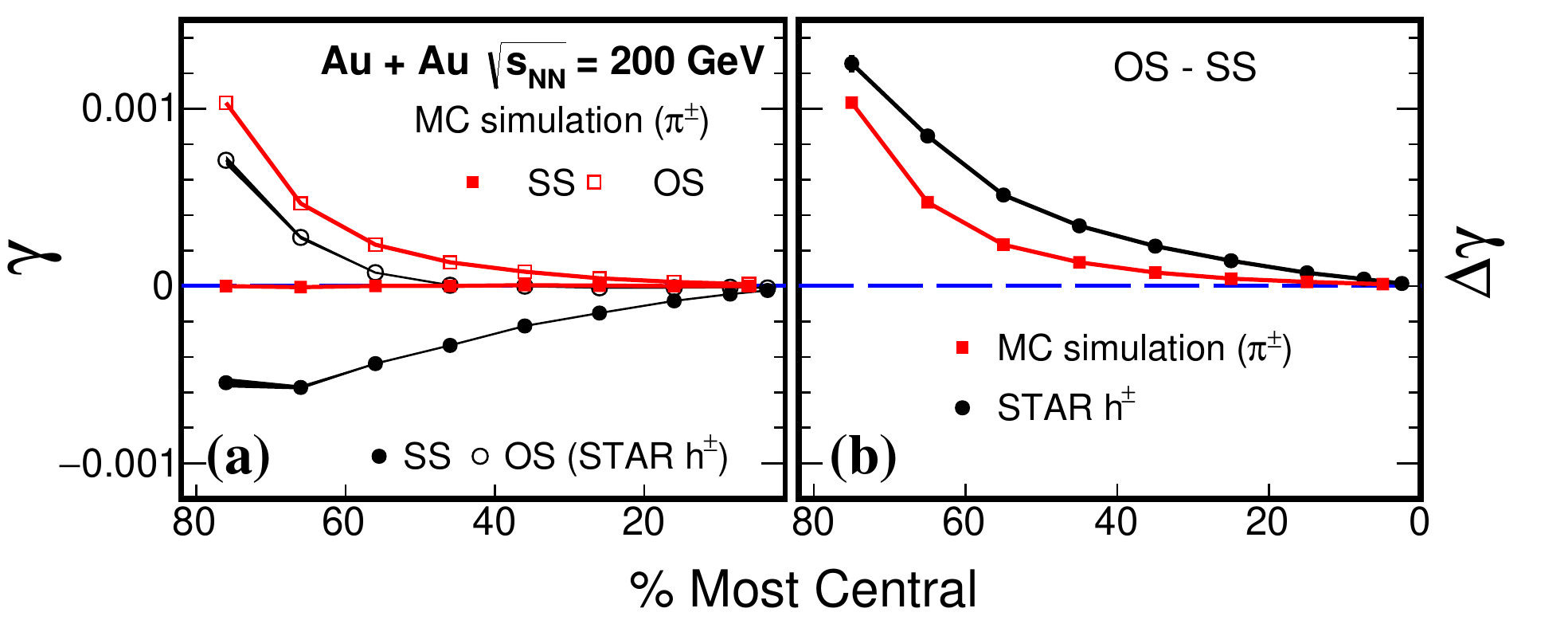}
\caption[]{(Color online) (a) $\gamma=\langle\cos(\alpha+\beta-2\psiPP)\rangle$ for same-sign ($\gSS$) and opposite-sign pairs ($\gOS$) within $|\eta|<1$ and (b) $\gdel=\gOS-\gSS$ vs.~centrality bin (1: most peripheral, 9: most central) from the resonance simulation ($\rho,\eta,\omega$), compared to STAR data from Ref.~\cite{Abelev:2009ac,Abelev:2009ad}.} 
\label{correlator_raw}
\end{figure}

{\it Simulation results and discussions.}
The correlator of Eq.~(\ref{eq:C}) is calculated with the cumulant method ~\cite{Bilandzic:2010jr}. 
We compute $\gamma$ w.r.t.~both $\psiPP$ (known from the simulation) and $\psiEP$ (reconstructed from random subevents and corrected for the EP resolution). 
The two results are consistent. 
Figure~\ref{correlator_raw}(a) shows $\gSS$ and $\gOS$
vs.~centrality from the simulation. 
The $\gSS$ is zero as expected because resonance decays do not affect SS particles.
The non-zero $\gOS$ in this simulation is due to correlation between the decay pions coupled with the resonance $v_{2}$.
For reference the STAR data of charged hadrons~\cite{Abelev:2009ac,Abelev:2009ad} are also displayed in Fig.~\ref{correlator_raw}(a)
\footnote{Note that the STAR data are not corrected for the \pT-dependent efficiency. We estimate that efficiency correction would reduce the data magnitude by 20\%.}.
The purpose of our study is not to compare simulation to data for the individual $\gOS$ and $\gSS$, but their difference $\gdel$. This is shown in Fig.~\ref{correlator_raw}(b). 
The simulation results are not far away from data. 
We note that in the MEVSIM model studied in the STAR publications~\cite{Abelev:2009ac,Abelev:2009ad} can also approximately describe the measured $\Delta\gamma$. 
The estimate by Voloshin~\cite{Voloshin:2004vk} which claims negligible contributions from resonance decays, however, appears to have missed a factor of $v_{2}$. 
Our results in Fig.~\ref{correlator_raw}(b) implies that the charge separation effect seen in data may come largely from resonance decays. 
About half of the signal strength comes from $\rho$ decays, and the other half come from $\eta$ and $\omega$ decays. 
This is because the three-body decays are similar to two-body decays in terms of the angular correlation strength $\mean{\cos(\alpha+\beta-2\phires)}$.

The finite $\gdel$ in the resonance simulation is due to correlation between the decay pions coupled to the resonance $v_2$. In order to ``eliminate'' this background, one resorts to $\vobs=0$ ~\cite{Adamczyk:2013kcb}. As discussed in the introduction, due to various correlation effects, this background may not be completely eliminated by $\vobs=0$. To elucidate this point, let us take a detour to a simple case study: only $\rho$ with fixed $\pt$ and fixed $\vrho$.
We compute $\gdel$ as a function of $\vrhoexe=\mean{\cos2(\phirho-\psiPP)}$, 
and find a linear dependence with vanishing intercept. 
We can obtain this background by the mixed-event technique, calculating the $\gamma$ and $\vrhoexe$ using the $\psiPP$ from another event. 
We find the same proportionality; the mixed-event faithfully describes the same-event background.
By subtracting the mixed-event $\gdel$ background, we obtain the real signal, and in this case the real signal is zero as it should be. 
\begin{figure}\centering
\includegraphics[width=0.49\textwidth]{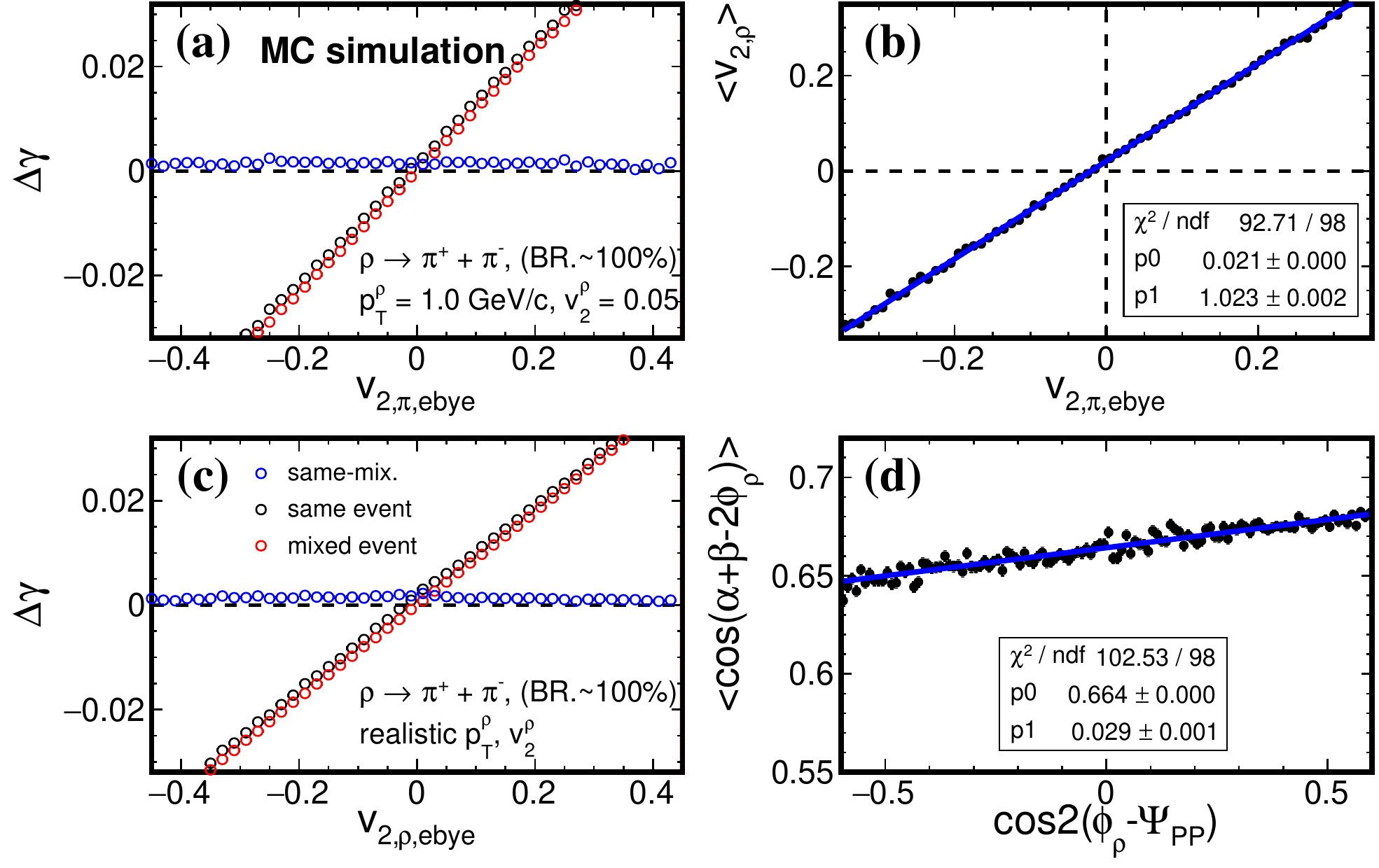}
\caption[]{(Color online) (a,b) Simulation of $\rho$ only, with fixed $\ptrho$ = 1.0~\GeVc and fixed $\vrho = 5\%$. (a) $\langle\cos(\alpha+\beta-2\psiPP)\rangle$ vs.~$\vpiobs$, and (b) $\mean{\vrho}$ vs.~$\vpiexe$; the finite $\mean{\vrho}$ at $\vpiexe=0$ is the reason why flow background cannot be removed completely by mixed-events or by $\vpiexe=0$. (c,d) Simulation of $\rho$ only, but with realistic $\ptrho$ and $\vrho$ distributions from 200~GeV Au+Au data. (c) $\langle\cos(\alpha+\beta-2\psiPP)\rangle$ vs.~$\vrhoobs$, and (d) correlation between decay angle $\mean{\cos(\alpha+\beta-2\phirho)}$ and $\cos2(\phi_{\rho}-\psiPP)$, induced by their dependencies on $\pt$. The correlation breaks the factorization in Eq.~(\ref{eq:C}) and is the reason why residual flow background still exists after mixed-event subtraction or by $\vrhoexe=0$. Only OS correlators are plotted; the SS correlators are all zero.} 
\label{correlator_v2}
\end{figure}

Unfortunately, it is very challenging, if not at all impossible, to measure $\vrhoexe$. One can only measure $\vexe$ of final-state particles, i.e.~all charged pions in the case of our simulation. 
Figure~\ref{correlator_v2}(a) shows $\gdel$ as a function of $\vpiexe=\mean{\cos2(\phi_{\pi}-\psiPP)}$ in same event (black points) and mixed event (red points). The same proportionality is observed for mixed-event, but the same-event result shows only an approximate proportionality; there is a finite intercept at $\vpiexe=0$. As a result, the mixed-event subtracted result (blue points) shows a finite $\gdel$. This is because, when $\vpiexe=0$, the average $\vrhoexe$ is positive 
as shown in Fig.~\ref{correlator_v2}(b). In this case the mixed event does not faithfully reproduce the background in the same event. There is remaining background even at $\vpiexe=0$.

Still generate only $\rho$ in the simulation but with realistic $\pt$ distribution and $\pt$-dependent $\vrho(\pt$). Figure~\ref{correlator_v2}(c) shows $\gdel$ as a function of $\vrhoexe$ for same event and mixed event. The same-event intercept is not exactly zero in this case, even with $\vrhoexe=0$. This is due to the induced correlation between the $\rho$ decay angle and $\vrho$, because both depend on $\pt$, as shown in Fig.~\ref{correlator_v2}(d). As a result the factorization in Eq.~(\ref{eq:bkgd}) does not strictly hold, and residual correlation remains even at $\vrhoexe=0$. For mixed events, on the other hand, the factorization is still valid because no such correlation exists. This implies that, even if $\vrhoexe$ can be experimentally assessed, one may be still unable to completely eliminate the flow background.

\begin{figure}\centering
\includegraphics[width=0.3\textwidth]{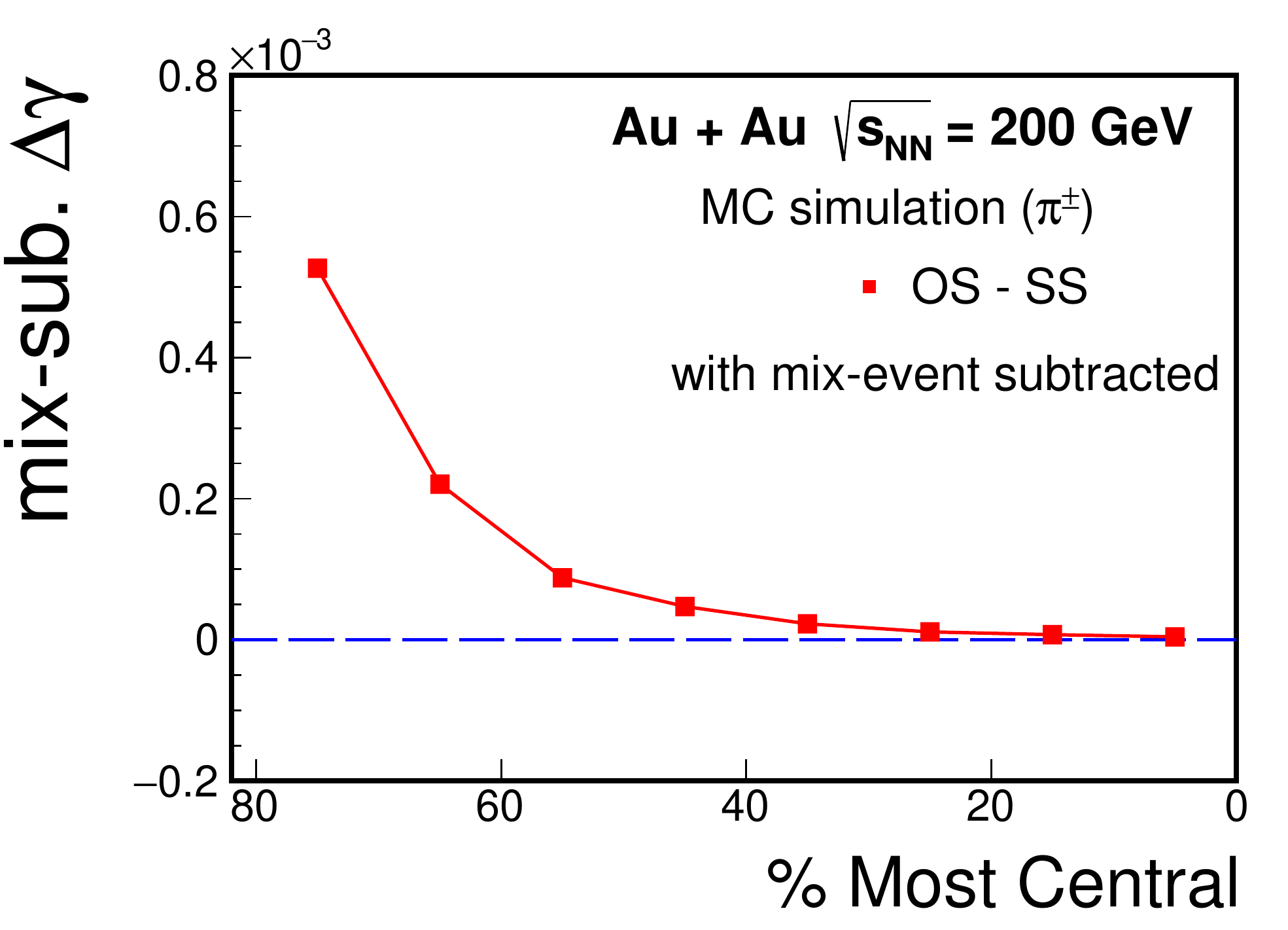}
\caption[]{(Color online) Prediction of the mixed-event subtracted $\gdel=\gOS-\gSS$ within $|\eta|<1$ 
vs.~centrality by the resonance simulation ($\rho,\eta,\omega$).
} 
\label{bkg_correlator}
\end{figure}

With the realistic resonance kinematic distributions, 
the situation using $\vpiexe$ to do analysis is, of course, worse; a significant residual background remains after mixed-event subtraction. 
It is, however, still the experimentally most viable way so far to suppress flow background.
%
Back to our realistic resonance simulation, 
Fig.~\ref{bkg_correlator} shows the prediction of the mixed-event subtracted $\gdel$ vs.~centrality, analyzed as a function of $\vpiexe$. 
It is reduced by approximately a factor of two from the inclusive signal in Fig.~\ref{correlator_raw}(b). In other words, the background is reduced by one half.


\begin{figure}\centering
\includegraphics[width=0.49\textwidth]{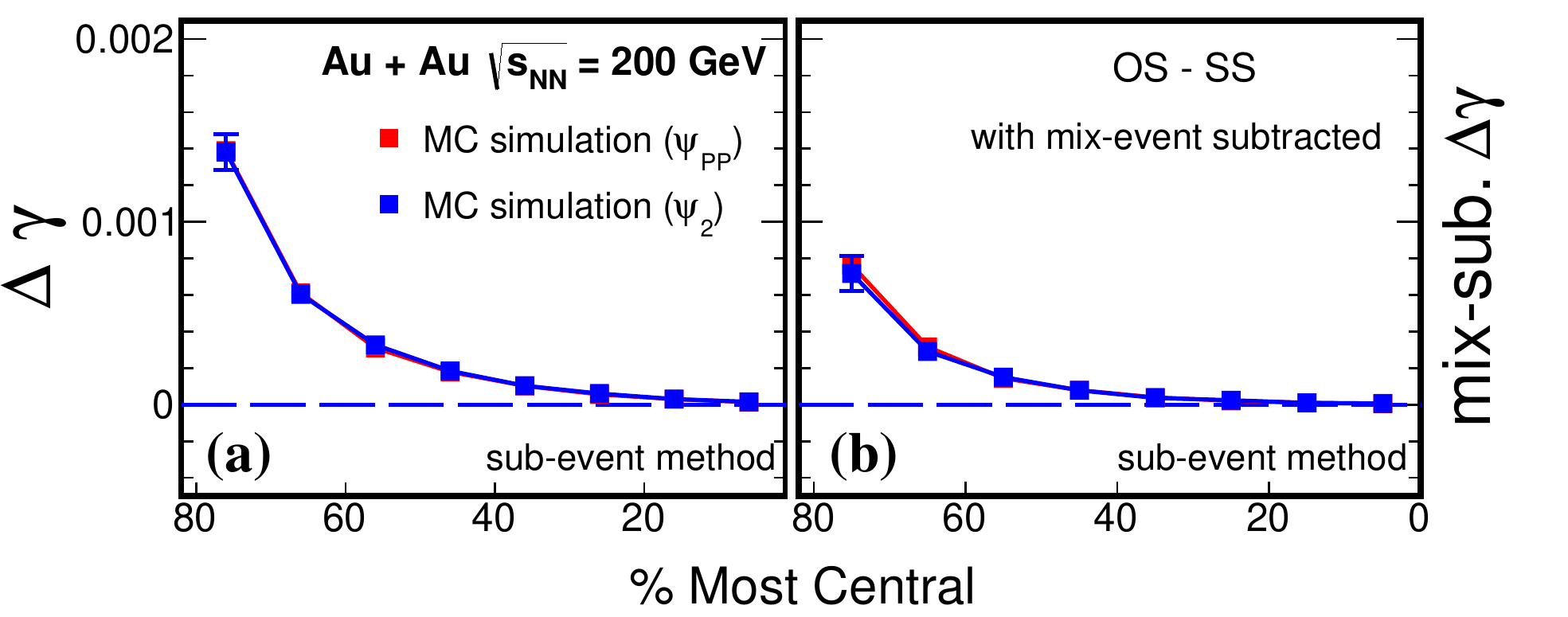}
\caption[]{(Color online) Predictions of $\langle\cos(\alpha+\beta-2\psi)\rangle$ within subevents (i.e.~averaged between $0<\eta<1$ and $-1<\eta<0$) vs.~centrality by the resonance simulation ($\rho,\eta,\omega$), using both $\psiPP$ and $\psiEP$ (reconstructed from the other subevent, i.e.~$-1<\eta<-0.05$ or $0.05<\eta<1$, and corrected for EP resolution). (a) Inclusive results, and (b) flow background suppressed results by mixed-event subtraction.} 
\label{bkg_correlator_sub}
\end{figure}

We have so far concentrated on the range $|\eta|<1$ (a.k.a.~STAR acceptance) and used directly the $\psiPP$. 
In the rest, we divide each event into two subevents: $-1<\eta<-0$ and $0<\eta<1$, and calculate $\gamma$ and $\vobs$ using one subevent with respect to the $\psiEP$ from the other subevent, with $0.05<\eta<1$ and $-1<\eta<-0.05$ respectively. 
Figure~\ref{bkg_correlator_sub}(a) show $\gdel$ vs.~centrality; the $\gdel$ has been corrected for EP resolution (obtained from sub-subevents within the subevent). 
For comparison, we also show the result by directly using $\psiPP$.
We note there can be difference between these two results due to interplay of two effects: an overestimated EP resolution because of correlations caused by decay daughters across the two 
sub-subevents, and correlations, due to the same reason, between particles for correlator calculation and those for EP reconstruction.

Figure~\ref{bkg_correlator_sub}(b) shows the $\gdel$ result after mixed-event subtraction. The mixed-event subtraction is performed as a function of $\vexe$ for the $\psiPP$ case and $\vobs$ for the $\psiEP$ case. For the latter, the mixed-event subtracted $\gdel$ is integrated over all events and then corrected for the EP resolution. Compared to panel (a), the flow background is reduced by approximately a factor of two, but not completely eliminated. These results serve as our predictions within subevent acceptance with and without flow background suppression.


{\it Summary.}
We have investigated the effect of resonance decays with a simple MC simulation on the charge separation observable $\gdel$, the three-particle correlator difference between opposite- and same-sign pairs used in search for the CME. We use realistic kinematic distributions and elliptic flow for the resonances, guided by data. 
We find that the simulation can largely account for the measured $\gdel$ by STAR. We also extract the $\gdel(\vobs=0)$, 
similarly to data analysis, to suppress the elliptic flow background. 
It is found that the flow background is reduced by approximately a factor of two, but not completely removed. 
This is because of two reasons: (i) the $\vobs$ is measured by final-state particles, not that of the resonances, 
whose zero value would eliminate the background; and (ii) even with zero $\vexe$ of the resonances, 
residual background remains due to the induced correlations between $\pt$-dependent flow and $\pt$-dependent decay angle. 
We make predictions for $\gdel(\vobs=0)$, solely due to resonance decays, in Au+Au collisions at 200~GeV, 
with and with background suppression.

We have only studied a few resonance decay channels. There are other decays, such as $K^*\rightarrow K\pi$ and $\Delta\rightarrow p\pi$, that may also contribute appreciably and should be investigated in the future. Besides resonance decays, there are other background sources of particle correlations, e.g.~jet correlations~\cite{Wang:2009kd,Petersen:2010di} which can be studied with event generators like HIJING~\cite{Gyulassy:1994ew}. We postpone such studies to future work.

In order to isolate CME from elliptic flow driven backgrounds, the collisions of
isobaric nuclei, $^{96}_{44}$Ru +$^{96}_{44}$Ru and $^{96}_{40}$Zr +$^{96}_{40}$Zr, have been proposed~\cite{Voloshin:2010ut},
The different magnetic fields in these collisions will result in roughly $10\%$ difference in the expected CME signal strengths, 
while the backgrounds due to different deformations of the isobaric nuclei are expected to differ less than $2\%$, for the centrality range of $20-60\%$ at \sNN = 200 GeV~\cite{Deng:2016knn}.
The isospin difference between the isobaric nuclei, while having no effect on $\rho$, $\eta$, $\omega$ production, 
is expected to affect the relative abundances of $\Delta^{++}, \Delta^{0}, K^{*}(892)$ whose decays can cause backgrounds to the $\gamma$ correlator measurements in addition to those backgrounds in Ref~\cite{Deng:2016knn}. 
The magnitudes of such effects are quantitatively unclear, and we plan to follow up on this in a future work.

The inputs to our simulation are from experimental measurements of resonances; these resonances must exist in the final state of real heavy ion collisions, so our simulation results, with the simplest physics involved, should have high relevance. 
Our study indicates that background removal from CME-related correlation measurements is more complicated than initially thought. Given the many sources of resonance decays and many of them not experimentally measured, it seems rather pessimistic to identify true CME signals using the three-particle correlator. 
Many-particle correlations should eliminate most of the resonance decay contributions. Whether many-particle correlations are useful or not to identify CME depends on the relative strengths of CME and other multiparticle correlations, such as jets.

{\it Acknowledgments.}
We thank Drs. Adam Bzdak, Berndt Mueller, and Gang Wang for helpful discussions. 
This work was supported by the U.S.~Department of Energy (Grant No.~DE-FG02-88ER40412).


\bibliography{ref}

\begin{thebibliography}{10}
\bibitem{PPRP:2017}
$\psi_{\rm PP}$ here refers to the plane span by the beam direction and the
  overlap geometry 2-nd harmonic symmetry plane. Due to fluctuations the
  $\psi_{\rm PP}$ is not necessarily the same as the reaction plane.
\end{thebibliography}
\end{document}